\documentclass[a4paper,11pt]{article}
\RequirePackage{plautopatch}
\usepackage{bm}
\usepackage{graphicx}
\usepackage{ascmac}
\usepackage{amsfonts}
\usepackage{amsmath,amssymb}
\usepackage{cancel}
\usepackage{url}
\usepackage{mathtools}
\usepackage{physics}
\usepackage{physics2}
\usephysicsmodule{braket}
\usephysicsmodule{ab}
\usepackage{derivative}
\allowdisplaybreaks[1]
\usepackage{pos}
\title{Properties of $X(3872)$ from hadronic potentials coupled to quarks}

\author*[a]{Ibuki Terashima}
\author[a]{Tetsuo Hyodo}

\affiliation[a]{Department of Physics, Tokyo Metropolitan University,\\
  1-1 Minami-Osawa, Hachioji-city, Tokyo 192-0397, Japan}
\emailAdd{terashima-ibuki@ed.tmu.ac.jp}
\emailAdd{hyodo@tmu.ac.jp}

\abstract{
The concept of compositeness is used to quantitatively discuss the hadronic molecular nature in exotic hadrons. In this work, we develop a formulation to explicitly introduce the compact quark state contribution in the hadronic potentials together with the direct four-point interaction. We derive analytic expressions of the compositeness from the effective potential between hadrons when the system has a bound state. Applying this formulation to the $X(3872)$, we examine the variation of the compositeness with respect to the binding energy, energy of $\chi_{c1}(2P)$, cutoff momentum, and strength of the direct $D^0\bar{D}^{*0}$ interaction.
}

\FullConference{The XVIth Quark Confinement and the Hadron Spectrum Conference (QCHSC24)\\
 19-24 August, 2024\\
 Cairns Convention Centre, Cairns, Queensland, Australia\\}

\begin{document}
\maketitle

\section{Introduction}

After the first observation of an exotic hadron candidate $X(3872)$ (also known as $\chi_{c1}(3872)$~\cite{ParticleDataGroup:2024cfk}) in 2003~\cite{Belle:2003nnu}, it remains an ongoing issue to uncover the internal structure of exotic hadrons. In contrast to ordinary hadrons, the $X(3872)$ is an extremely shallow bound state with the binding energy $B = 40$ keV measured from the $D^0\bar{D}^{*0}$ threshold. The $X(3872)$ is a charmonium-like state possessing the quantum numbers $J^{PC}=1^{++}$. However, its mass is substantially smaller than the $\chi_{c1}(2P)$ state predicted in the constituent quark model~\cite{Godfrey:1985xj}, indicating an exotic internal structure beyond the simple $c\bar{c}$. Regarding the internal structure of the $X(3872)$, there are two major candidates: a compact multiquark state involving the mixing of $c\bar{c}u\bar{u}$ and $c\bar{c}d\bar{d}$ components~\cite{Maiani:2004vq}, and a molecular state of $D^0\bar{D}^{*0}$~\cite{Close:2003sg,Gamermann:2009uq} suggested by its very narrow decay width. Among others, the most plausible scenario is a $D^0\bar{D}^{*0}$ molecule dominant state mixed with a non-negligible $\chi_{c1}(2P)$ component~\cite{Takizawa:2012hy,Hosaka:2016pey}.

To describe the $X(3872)$, we have developed a theoretical framework where the $D^0\bar{D}^{*0}$ is induced by the coupling to the compact $c\bar{c}$ state~\cite{Terashima:2023tun}. Applying the Feshbach method to the coupled-channel system of quark and hadron degrees of freedom, we obtain a non-local effective $D^0\bar{D}^{*0}$ potential including the $c\bar{c}$ contributions. With the use of the Yukawa-type transition form factor, the effective potential and the scattering phase shifts are analytically derived. In this model, the $X(3872)$ is expressed by the superposition of the $D^0\bar{D}^{*0}$ molecule and the compact $c\bar{c}$ state.

The concept of compositeness is used as a method to quantitatively characterize the molecular component in bound states. Compositeness was first introduced by S. Weinberg in a relation connecting the scattering length and effective range to the compositeness of $s$-wave shallow bound states~\cite{Weinberg:1962hj,Weinberg:1965zz}. In the case of the $X(3872)$, the composite component represents $D^0\bar{D}^{*0}$ molecule, and the elementary component corresponds to the compact $c\bar{c}$ state. In this context, the compositeness $X$ is defined as the probability of finding the molecular component in the bound state, which takes a value in the range $0\leq X \leq 1$. Recently, the compositeness has been applied to study the internal structure of exotic hadrons, in particular the near-threshold states~\cite{Baru:2003qq,Hyodo:2011qc,Hyodo:2013nka,Oller:2017alp,Guo:2017jvc,vanKolck:2022lqz,Kinugawa:2024crb}.

In this work, we generalize the framework of Ref.~\cite{Terashima:2023tun} by explicitly introducing the four-point direct interaction of $D^0\bar{D}^{*0}$, and evaluate the compositeness of the $X(3872)$. We show that the compositeness $X$ is also analytically obtained by using the bound state wave function. We examine the parameter dependence of the compositeness $X$ by varying the binding energy, energy of the $c\bar{c}$ state, momentum cutoff, and strength of the direct hadron interaction. Similar disucussion can be found in Ref.~\cite{Song:2023pdq}.

\section{Formulation}
\subsection{Channel coupling}

Adopting basically the same notation with Ref.~\cite{Terashima:2023tun}, we start from the channel-coupled Hamiltonian $H$:
\begin{align}\label{eq:hamiltonian}
        H=
    \begin{pmatrix}
        T^{c\bar{c}} & 0             \\
        0     & T^{D^0\bar{D}^{*0}}+ \Delta
    \end{pmatrix}
    +
    \begin{pmatrix}
        V^{c\bar{c}} & V^t   \\
        V^t   & V^{D^0\bar{D}^{*0}}
    \end{pmatrix},
\end{align}
where $T^{c\bar{c}}$ and $T^{D^0\bar{D}^{*0}}$ are the kinetic energies of the $c\bar{c}$ and $D^0\bar{D}^{*0}$ channels, $\Delta$ is the threshold energy of $D^0\bar{D}^{*0}$, $V^{c\bar{c}}$ and $V^t$ are the confinement potential of $c\bar{c}$ and the transition potential between $c\bar{c}$ and $D^0\bar{D}^{*0}$. In addition to those considered in Ref.~\cite{Terashima:2023tun}, we newly introduce the direct hadron channel potential $V^{D^0\bar{D}^{*0}}$. The Schr\"odinger equation reads 
\begin{align}\label{eq_2ch_schrodinger}
    H \ket{\Psi} =  E \ket{\Psi},  \quad  \ket{\Psi}=
    \begin{pmatrix}
        \ket{c\bar{c}} \\
        \ket{D^0\bar{D}^{*0}}
    \end{pmatrix},
\end{align}
where $\ket{c\bar{c}}$ and $\ket{D^0\bar{D}^{*0}}$ represent the wave function of the $c\bar{c}$ and $D^0\bar{D}^{*0}$ components.

Based on the Feshbach's method \cite{Feshbach:1958nx,Feshbach:1962ut}, the effective $D^0\bar{D}^{*0}$ interaction including the $c\bar{c}$ coupling contribution can be derived from the Hamiltonian~\eqref{eq:hamiltonian}. The coordinate space representation of the effective interaction is given by
\begin{align}
    V(\bm{r'},\bm{r},E)
        =\braket[3]{\bm{r'}}{V^h}{\bm{r}} + \frac{ \braket[3]{\bm{r'}}{V^t}{\phi_{0}} \braket[3]{\phi_0} {V^t}{\bm{r} }}{E-E_0}. \label{eq_VDD}
\end{align}
Here we take into account only the intermediate $\chi_{c1}(2P)$ state denoted by $\ket{\phi_0}$ as an eigenstate of $T^{c\bar{c}}+V^{c\bar{c}}$ which gives the most dominant contribution to the $s$-wave $D^0\bar{D}^{*0}$ interaction. In the following, we consider the Yukawa type transition form factor $\braket[3]{\phi_0}{V^t}{\bm{r}}=g_0e^{-\mu r}/r$ with the coupling constant $g_0$ and the momentum cutoff $\mu$. By assuming the separable $D^0\bar{D}^{*0}$ interaction with the same form factors, $\braket[3]{\bm{r}^\prime}{V^h}{\bm{r}}=\omega^h(e^{-\mu r^\prime}/r^\prime)( e^{-\mu r}/r)$, the effective $D^0\bar{D}^{*0}$ interaction $V_\mathrm{eff}^{h}(\boldsymbol{r'}, \boldsymbol{r}, E)$ is expressed as
\begin{align}\label{eq_VDDeff_omega_h}
    V(\boldsymbol{r'}, \boldsymbol{r}, E) = 
    \omega(E) \frac{e^{-\mu r}}{r} \frac{e^{-\mu r'}}{r'},
    \quad\omega(E) = \omega^h +  \omega^q(E),
    \quad
     \omega^q(E)=  \frac{g_0^2}{E - E_0} .
\end{align}
By solving the Schr\"odinger equation for this non-local energy-dependent potential with an appropriate boundary condition, we obtain the analytic expression of the bound state wave function~\cite{Aoki:2021ahj}
\begin{align}\label{eq:psi_yukawa}
    \Psi_{E=-B} (r)  \propto \pab{ -\frac{\kappa e^{-\kappa r}}{r} + \frac{\kappa e^{- \mu r }}{r} } ,
\end{align}
where $\kappa=\sqrt{2m B}$ with the binding energy $B$. The normalization of the wave function for an energy dependent interaction requires special care, which we discuss in section~\ref{subsec:compositeness}.

The binding energy $B$ is determined by the bound state condition. For this purpose, we utilize the Lippmann-Schwinger equation for the T-matrix 
\begin{align}\label{eq:LSeq}
    T(E)    &= \frac{1}{\frac{1}{\omega(E)}+I(E)}, 
\end{align}
where the function $I(E)$ is defined as 
\begin{align}
        I(E)    &=-\int \frac{d^3 {\bm k}}{(2\pi)^3} \frac{1}{E-k^2/(2m)+i\epsilon}\pab{\frac{4\pi}{\mu^2+k^2} }^2 
        = \frac{4\pi m}{\mu\pab{\sqrt{-2mE-i\epsilon}+\mu}^2} .\label{eq:I_sekibun} 
\end{align}
Because the bound state is expressed by the pole of the T-matrix, the bound state condition is given by 
\begin{align}
        \frac{1}{\omega(-B)}+I(-B) &=0 . \label{eq:pole_cndtn}
\end{align}

For a given binding energy $B$, the strength of the direct $D^0\bar{D}^{*0}$ interaction is constrained by the Hermiticity of the Hamiltonian, namely the condition $g_0^2\geq 0$. From Eq.~\eqref{eq:pole_cndtn}, the coupling constant can be expressed as
\begin{align}
    g_0^2 &= \left(B + E_0\right) \left(\frac{1}{I(-B)} + \omega^h\right) .
\end{align}
By choosing the bare energy $E_0>-B$, we obtain the condition 
\begin{align}\label{eq:omega_h_lim}
    \omega^h \geq \omega^h_\mathrm{b},\quad\omega^h_\mathrm{b}=-\frac{1}{I(-B)}.
\end{align}
Because Eq.~\eqref{eq:I_sekibun} indicates $I(-B)>0$, the lower bound $\omega^h_\mathrm{b}$ is negative. At $\omega^h=\omega^h_{\rm b}$, the coupling constant vanishes, $g_0^2=0$. In other words, the bound state is generated purely by the direct $D^0\bar{D}^{*0}$ interaction, indicating the molecular (composite) nature of the bound state. There is no upper bound on $\omega^h$, because the repulsive direct interaction can always be compensated by the attractive $c\bar{c}$ coupling term so as to obtain the suitable attraction for the bound state.

\subsection{Compositeness}\label{subsec:compositeness}

Let us calculate the compositeness of the bound state obtained in the above formulation with two different methods. First, we use the normalization condition of the bound state wave function.
The normalization of the wave function with an energy-dependent non-local potential is given by~\cite{Miyahara:2018onh}
\begin{align}
    1 = \left. \int d^3{\bm{r'}} d^3{\bm{r}}  \Psi^*_{E}(\bm{r'}) 
    \left[
      \delta(\bm{r'} -\bm{r})-  \pdv{V(\bm{r'},\bm{r},E)}{E} 
    \right]
    \Psi_E(\bm{r}) \right|_{E=-B},
    \label{eq:normalization}
\end{align}
and the compositeness $X$ and elementarity $Z$ are respectively defined as
\begin{align}
    X=\int d^3{\bm{r}} |\Psi_{E=-B}(\bm{r}) |^2 ,\quad
    Z= \left. -\int d^3{\bm{r'}} d^3{\bm{r}}  \Psi^*_{E}(\bm{r'}) \pdv{V(\bm{r'},\bm{r},E)}{E} \Psi_E(\bm{r}) \right|_{E=-B} ,
    \label{eq:Xwf}
\end{align}
with
\begin{align}
    X+Z=1 .
\end{align}
Normalizing the wave function~\eqref{eq:psi_yukawa} by Eq.~\eqref{eq:normalization}, we obtain the analytic expression of the compositeness by Eq.~\eqref{eq:Xwf}
\begin{align}
    X &= \left(1 + \frac{g_0^2 \kappa \mu (\kappa + \mu)^3}{8\pi m^2 (g_0^2 - (B + E_0)\omega^h)^2}\right)^{-1} .\label{eq:X_bound}
\end{align}

Next, we evaluate the compositeness from the Lippmann-Schwinger equation~\eqref{eq:LSeq}. The compositeness can be expressed by the effective interaction and loop function as~\cite{Kamiya:2015aea,Kamiya:2016oao}
\begin{align}
    X &= \frac{-I'(E)}{- I'(E)-(\omega^{-1})' } \bigg|_{E = -B} ,
    \quad
    Z = \frac{-(\omega^{-1})'}{- I'(E)-(\omega^{-1})' } \bigg|_{E = -B}.
\end{align}
where $\prime$ represents the energy derivative. 
Substituting Eqs.~\eqref{eq_VDDeff_omega_h} and \eqref{eq:I_sekibun}, we obtain
\begin{align}
    X &= \left(1 + 2\pi \frac{g_0^2}{(B + E_0)^2} \frac{\kappa}{\mu (\mu + \kappa)}\right)^{-1} .
\end{align}
It is possible to show that this is is equivalent to the expression~\eqref{eq:Xwf} with the help of the bound state condition~\eqref{eq:pole_cndtn}. 

\section{Numerical results}

For numerical calculations, we use the parameters based on Ref.~\cite{Terashima:2023tun} (see Table~\ref{tab:param_X3872}). We choose the binding energy of the $X(3872)$ ($B$) as the central value in PDG~\cite{ParticleDataGroup:2024cfk}. The bare energy of $\chi_{c1}(2P)$ ($E_0$) is taken from the constituent quark model~\cite{Godfrey:1985xj}. As in Ref.~\cite{Terashima:2023tun}, we choose the cutoff parameter $\mu$ as the pion mass, and set $\omega^h=0$. In this case, the value of the compositeness is obtained as $X=0.99$. This almost complete molecular nature of the $X(3872)$ is a consequence of the extremely small binding energy, $B=40$ keV. 

\begin{table}
    \centering
    \begin{tabular}{cc}
         parameter & representative value \\ \hline
         binding energy $B$ & $40$ keV \\
         bare energy $E_0$ &  $78$ MeV \\
         cutoff $\mu$ &  $140$ MeV \\
         direct interaction $\omega^h$ & 0 \\
    \end{tabular}
    \caption{Representative values of the parameters for the $X(3872)$.}
    \label{tab:param_X3872}
\end{table}

\begin{figure}
    \centering
    \includegraphics[width=0.49\linewidth]{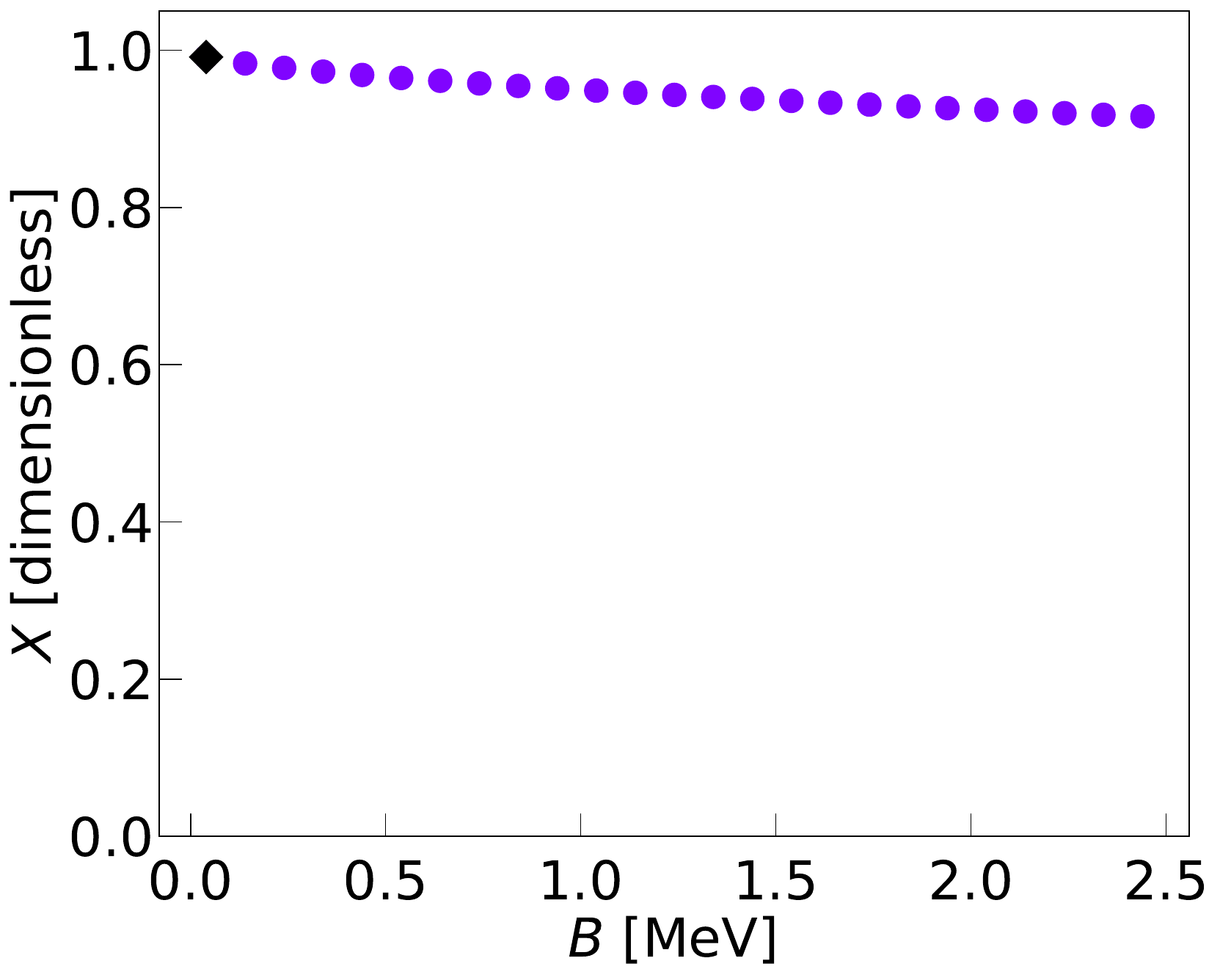}
    \includegraphics[width=0.49\linewidth]{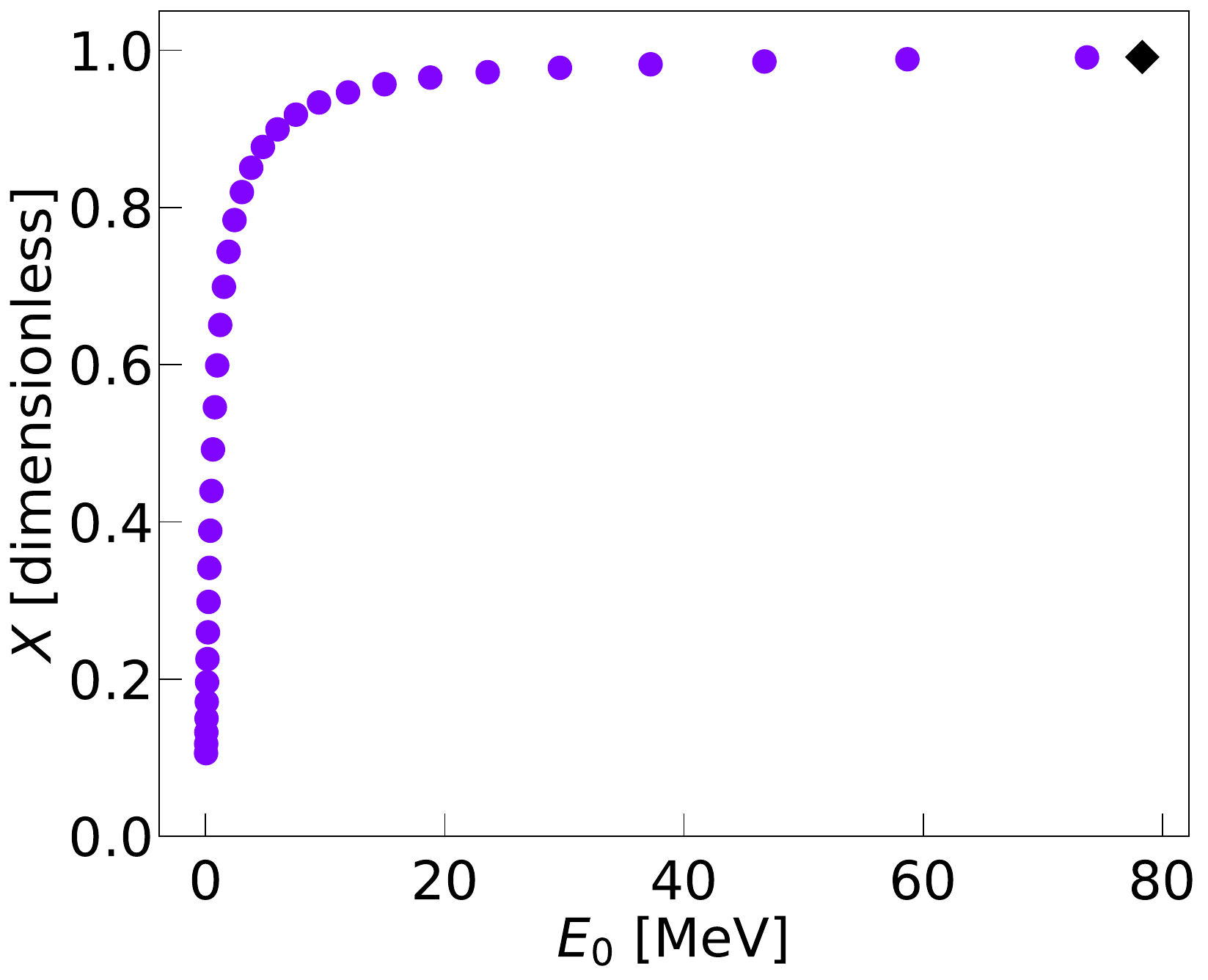}
    \caption{(Left) Compositteness $X$ as a function of binding energy $B$. (Right) Compositteness $X$ as a function of $\chi_{c1}(2P)$ energy $E_0$. The diamond symbols represent the result with the parameters in Table~\ref{tab:param_X3872}.}
    \label{fig:X_B-and-X_E0}
\end{figure}

Next, we vary one of the parameters with others fixed, in order to study the sensitivity of the compositeness to these parameters. We first study the dependence on the binding energy $B$. For this purpose, we fix $E_0$, $\mu$, $\omega^h$ as the values in Table~\ref{tab:param_X3872}, and vary the coupling constant $g_0$ to adjust $B$. In the left panel of Fig.~\ref{fig:X_B-and-X_E0}, we show the result of the compositeness $X$ in the range 40 keV $\leq B\leq $ 2.5 MeV where the upper bound is given by the singularity of the scattering amplitude from the separable Yukawa potential, $\kappa=\mu/2$. The result with $B=40$ keV is marked by the diamond symbol. This figure shows that the compositeness $X$ decreases as the binding energy $B$ increases. This is consistent with the weak binding limit $B\to 0$, where the bound state becomes a purely molecular state, $X \to 1$~\cite{Hyodo:2014bda,Hanhart:2014ssa,Kinugawa:2023fbf}. In the plotted range of this figure, the dependence of $X$ on $B$ is not significant, and the $X(3872)$ stays as a molecular dominant state. 

The dependence on the bare energy $E_0$ is shown in Fig.~\ref{fig:X_B-and-X_E0}, right. With $\mu$ and $\omega^h$ being fixed, we set $B=40$ keV, and investigate the region $-B\leq E_0\leq 78$ MeV by adjusting $g_0$ along with $E_0$. The diamond symbol represent the original result with $E_0=78$ MeV. From this figure, we find that the compositeness $X$ increases as the bare energy $E_0$ increases. This is because we need more self energy correction to obtain the given binding energy when $E_0$ is larger~\cite{Kinugawa:2023fbf}. As long as $E_0$ is sufficiently large, the compositeness is almost unity, as a consequence of the low-energy universality. However, when the bare energy $E_0$ becomes as small as the binding energy $B$, then the compositeness deviates from unity and finally $X$ vanishes for $E_0\to -B$. Namely, the compositeness of the shallow bound state, such as the $X(3872)$, is typically obtained as $X=1$, and a non-composite bound state can be realized only through the fine-tuning of the parameter $E_0$~\cite{Kinugawa:2023fbf}.

\begin{figure}
    \centering
    \includegraphics[width=0.49\linewidth]{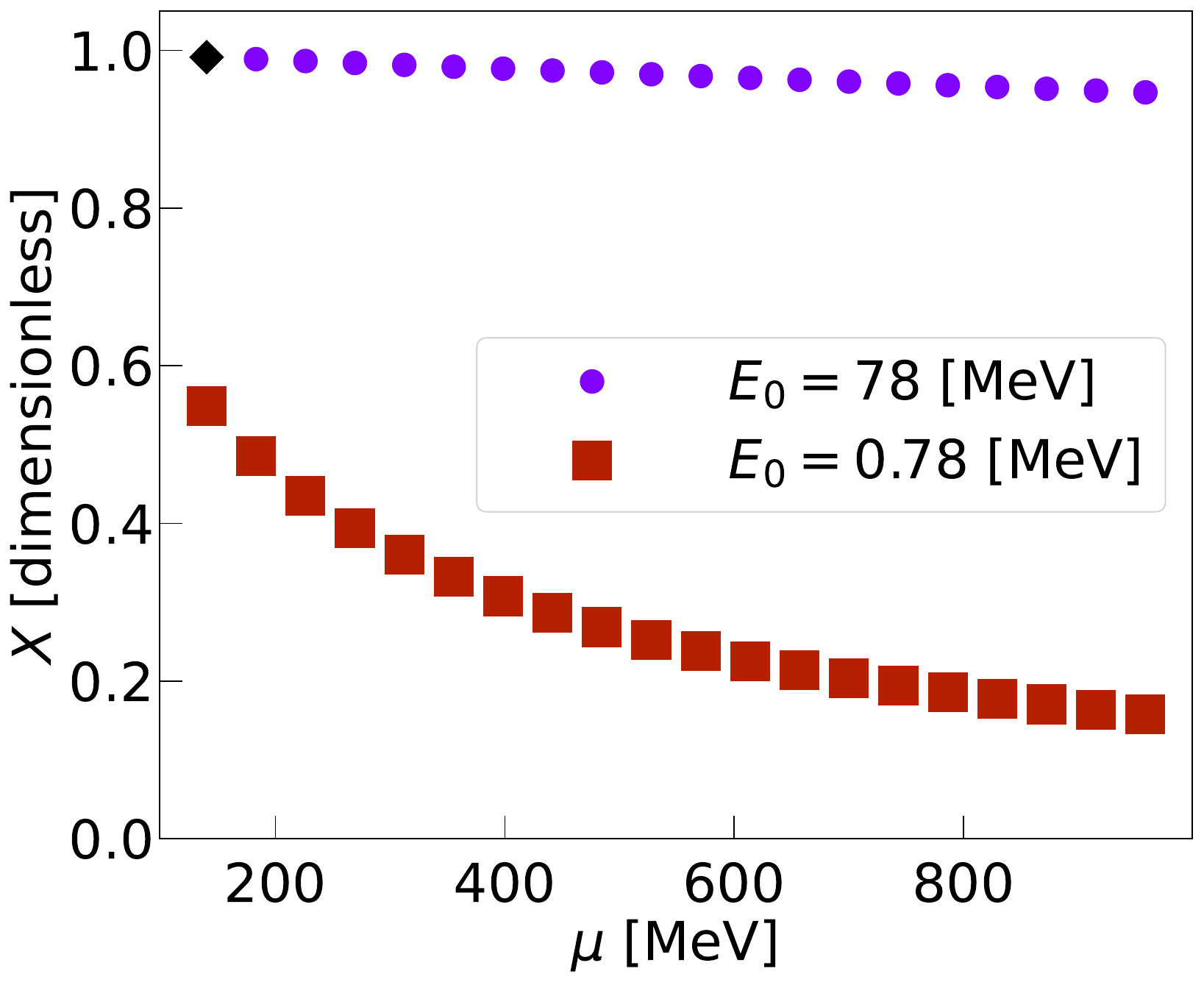}
    \includegraphics[width=0.49\linewidth]{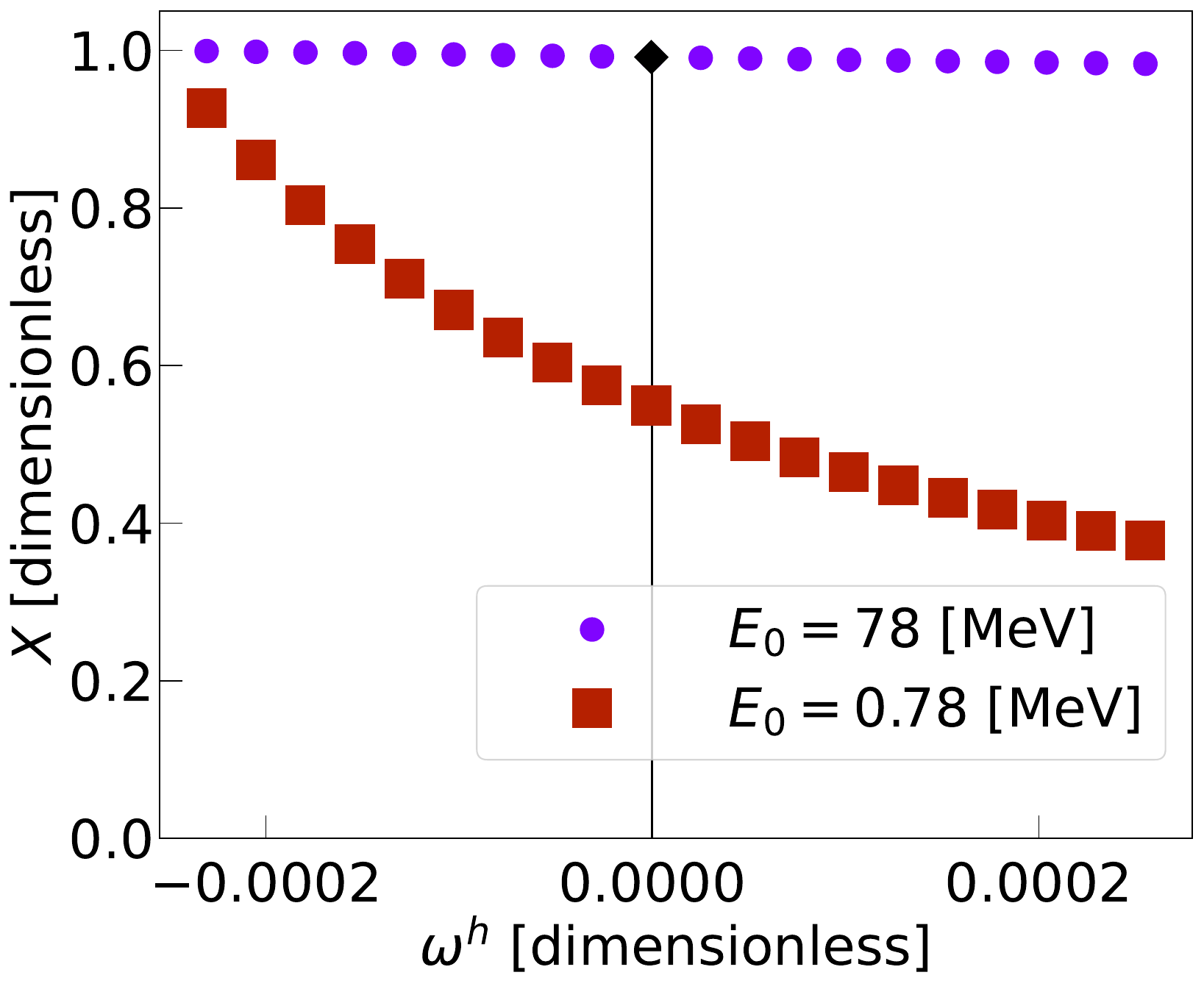}
    \caption{(left) Compositteness $X$ as functions of the cutoff $\mu$. (Right) Compositteness $X$ as functions of the direct interaction $\omega^h$. Circles (squares) stand for the results with $E_0=78$ MeV ($E_0=0.78$ MeV). The diamond symbols represent the result with the parameters in Table~\ref{tab:param_X3872}.}
    \label{fig:X_mu-and-omega_h}
\end{figure}

Figure~\ref{fig:X_mu-and-omega_h} (left) shows the compositeness $X$ as a function of $\mu$. Again, we fix $E_0$ and $\omega^h$ as in Table~\ref{tab:param_X3872}, and vary $g_0$ so as to reproduce $B=40$ keV for a given $\mu$. The cutoff is varied in the range 140 MeV $\leq \mu\leq$ 1 GeV. The results are plotted by the circles. For comparison, we show the results with $E_0=0.78$ MeV by the square symbols where the compositeness is about 0.5 at $\mu=140$ MeV. In both cases, the compositeness $X$ decreases as the cutoff $\mu$ increases. This feature is more prominent for the $E_0=0.78$ MeV case, where the compositeness is not close to unity at $\mu=140$ MeV. In other words, the compositeness of a shallow bound state is less sensitive to the variation of the cutoff parameter, due to the low-energy universality. 

Finally, we vary the strength of the direct $D^0\bar{D}^{*0}$ interaction $\omega^h$. We use $E_0$ and $\mu$ in Table~\ref{tab:param_X3872} and compensate for the variation of $\omega_h$ by the modification of $g_0$ so that the binding energy $B=40$ keV remains unchanged. We vary the parameter in the region $\omega^h_{\rm b}\leq \omega^h\leq -\omega^h_{\rm b}$, where the lower bound $\omega^h_{\rm b}<0$ is the attraction with which the bound state is generated purely by the direct $D^0\bar{D}^{*0}$ interaction, and the upper bound corresponds to the repulsion with the same interaction strength. In the right panel of Fig.~\ref{fig:X_mu-and-omega_h}, we show the compositeness $X$ with $E_0=78$ MeV ($E_0=0.78$ MeV) as a function of $\omega^h$ by the circles (squares). We find that the compositeness increases when the attractive direct interaction ($\omega_h<0$) is added. In contrast, the repulsive direct interaction ($\omega_h>0$) tends to decrease the compositeness. These results can be understood by considering that the interaction induced by the coupling to the $c\bar{c}$ state [$\omega^q(E)$ in Eq.~\eqref{eq_VDDeff_omega_h}] is always attractive, and the total attraction at $E=-B$ is fixed by the bound state condition. Thus, when an attractive $\omega^h$ is added, $\omega^q(E)$ needs to be decreased, and the molecular component induced by the direct interaction $\omega^h$ increases. As a result, the compositeness increases by adding the attractive $\omega^h$. On the other hand, repulsive $\omega^h$ requires stronger attraction from $\omega^q(E)$ to achieve the same binding energy, and the coupling to the $c\bar{c}$ state, which is the origin of the elementarity, increases~\cite{Kinugawa:2023fbf}. 

\section{Summary}

We have developed a model of the $X(3872)$ generated by the direct $D^0\bar{D}^{*0}$ interaction and the coupling to the $c\bar{c}$ state. In this model, the $X(3872)$ is expressed as the superposision of the $D^0\bar{D}^{*0}$ molecular component and the compact $c\bar{c}$ component. The expression of the $D^0\bar{D}^{*0}$ compositeness of the $X(3872)$  is obtained analytically. 

We examine the sensitivity of the compositeness of the $X(3872)$ to the binding energy, bare energy of the $\chi_{c1}(2P)$ state, cutoff, and strength of the direct $D^0\bar{D}^{*0}$ interaction. It is found that the compositeness decreases along with the increase of the binding energy, in accordance with the behavior of the weak-binding limit. The compositeness decreases with the reduction of the bare energy, which becomes prominent when the bare energy is fine-tuned down to the same order of magnitude with the binding energy. The dependence of the cutoff and the strength of the direct interaction is not strong when the original compositeness is close to unity, while the sizable parameter dependence arises for the bound state which is not dominated by the molecular component originally. 

As a future prospect, it is legitimate to examine the observable quantities such as the phase shift, scattering length, and effective range, as well as the behavior of the bound state wave function, in relation with the variation of the compositeness. 

\section*{Acknowledgements}
This work has been supported in part by the Grants-in-Aid for Scientific Research from JSPS (Grants
No.~JP23H05439, 
No.~JP22K03637, and 
No.~JP18H05402), 
and by JST, the establishment of university fellowships towards the creation of science technology innovation, Grant No. JPMJFS2139, 
by JST SPRING, Grant Number JPMJSP2156, 
and by the Sasakawa Scientific Research Grant from The Japan Science Society. 




\end{document}